\documentstyle[preprint,aps]{revtex}
\begin{document}
\draft     
\title
{$f_{N \pi N}$: from quarks to the pion derivative coupling} 
\author{P.  Bicudo and J. Ribeiro }
\address{
Departamento de Fisica and CFIF-Edificio Ciencia\\
 Instituto Superior Tecnico, Avenida Rovisco Pais,
1096 Lisboa Codex, Portugal}
\maketitle
\begin{abstract}
We study the $N \pi N$ coupling, in the framework of a QCD-inspired confining 
Nambu-Jona-Lasinio model.
A simple relativistic confining and instantaneous quark model is reviewed.
The Salpeter equation for the $qqq$ nucleon and the $q\bar{q}$ boosted
pion is solved.
The $f_{n\pi n}$ and $f_{n\pi \Delta}$ couplings are calculated  and they turn
 out to be reasonably good. 
The sensibility of $f_{n\pi n}$ and $f_{n\pi \Delta}$ to confinement, chiral 
symmetry breaking and Lorentz invariance is briefly discussed.
\end{abstract}

\section{INTRODUCTION}
It is widely accepted that the low energy phenomenology can be understood 
in terms of the mechanism of chiral symmetry spontaneous breaking. In this 
sense any microscopic theory (including Q.C.D.) which has a "correct" chiral 
limit will fare equally well in describing low energy hadronic phenomena. 
There is however an extra ingredient, the typical hadronic size, which  
should also play an important role in scattering. To see that this scale
is important it is sufficient to consider exotic reactions like K-n
scattering for this scale  controls the extent of overlap between the bare
nucleon and the kaon. In turn this overlap, when taken together with color
saturation and a generic spin-spin interaction for the quarks, accounts for
the generic hadronic central repulsion, a feature which is notorious in the
n-n system \cite{Ribeiro} and one which can also be measured in the K-n
exotic s-wave phase shifts \cite{KN}. Exotic reactions like $K-n$, where the 
$\overline{s}$ quark cannot annihilate with any of the quarks intervening in 
the reaction can be thought as an effective $K-n-K-n$ vertex without the 
exchange of $s$-channel resonances. Nor it has $t$-channel exchange of 
pions due to G-parity and therefore constitutes an ideal reaction to 
probe the low energy content of both the nucleon and the kaon wave functions, 
i.e. the overlap kernel.

On the other hand the bare rho-pion mass difference is completely controlled
by the extent of chiral symmetry breaking and this mechanism when understood
in terms of quarks, amounts to a specific connection between the chiral
condensate and the strength of the microscopic hyperfine interaction 
 $\vec s\ \cdot\ \vec s $ \cite{BicRib1,BicKre,VilLiu}.
It happens that the obtained strength of this $\vec s\ \cdot\ \vec s $ when
considered in the exotic K-n s-wave system yields good phase shifts provided
we have a small bare nucleon core. At this stage it should be understood that
the size of bare hadrons and the extent of the chiral condensate are not
independent quantities but instead they are related to each other. This is a  
consequence of the fact that in hadronic reactions the mechanism of 
spontaneous chiral symmetry breaking is self-consistently exerted in two 
separate sectors: in the Salpeter amplitudes of the intervening hadrons and 
in the  modifications it introduces in the quark-quark effective 
potentials.

It is therefore interesting to see if it is possible to describe in a 
unified way this central repulsion (linked with non-annihilating quark 
amplitudes) together with quark-antiquark annihilating amplitudes which 
are  not only at the origin of hadronic 
attraction, the other prominent feature of hadronic scattering and one which 
can be associated to n-n  peripheral attraction, but also at the origin of
other effects like strong decay rates.

The n-n peripheral attraction is a central issue at the crossroads between 
nuclear physics and the hadronic physics. Several attempts to describe this 
attraction in terms of Skyrmion Lagrangians have been made but the overall 
result remains inconclusive. The Skyrmion picture although being quite adapted in 
describing n-n repulsion in terms of topological numbers and despite 
containing the physics of the pion interaction, it is not simple to accommodate 
with the Yukawa picture of n-n attraction in terms of mesonic exchange and 
several mechanisms have been proposed in the literature to provide for this 
attraction\cite{Skyrme}. On the other hand the physics of the central n-n 
repulsion lies outside the domain of the chiral perturbation theory and 
therefore we cannot use this theory to describe the n-n scattering.

Any effective theory attempting to describe n-n attraction must also  be able  
to describe the physics of meson-nucleon systems among which we have 
the $\pi n$ system. In this respect the exotic 
K-n scattering is much simpler to study than the $\pi n$ reaction which 
being non-exotic, has now quark antiquark annihilating amplitudes which were 
absent in the exotic K-n. Also we are forced to consider the quasi-Goldstone 
nature of the pion which can be translated in the Salpeter language by the 
existence of two almost degenerate amplitudes (the so-called E-spin amplitudes)
 to describe the pion wave function in contrast with the kaon case which is 
well described by one Salpeter amplitude.
 
The Nambu, Jona-Lasinio effective lagrangian (N.J-L.)\cite{NJL} 
with a non-local potential\cite{BicRib1,BicKre,AmeYao,BicRib2,BicRib3,AdlDav} 
offers then a simple framework to describe in a unified way (in the 
present case with only one parameter for the potential strength besides 
the current quark masses) not only the bulk of hadronic spectroscopy but 
also hadronic scattering phenomena among which we have the coupled n-n,
the $\Delta -n$ and the $\Delta -\Delta$ scattering processes. It is clear  
that these processes, when seen from the quark microscopic point of view 
embodied in the N.J-L. lagrangian, should correspond to different scattering 
processes described by the $same$ set of lagrangian parameters. In this paper we 
will use N to represent either the $\Delta $ or the nucleon in the cases 
where it does not matter and reserve the letter n for the cases which are 
specific to the nucleon alone.

When dealing with potentials, Lorentz covariance becomes a problem not 
unrelated with the problem of relativistic description of bound states. 
Covariant generalizations of potentials are available in the literature 
\cite{Covar,Horvat} but at this stage we will ignore this issue which will be 
essential for quantitative predictions but certainly not for the qualitative 
picture which as we will show can be understood in terms of chiral symmetry 
and color confinement.

In nuclei the necessity of having a microscopic description of the nucleon in
terms of quarks is already felt when considering Coulomb and magnetic form
factors for s-shell and p-shell nuclei like $^6 Li$ and $^{16}O$ \cite{Spit}.
But the simple fact that chiral symmetry is spontaneously broken through the 
appearance of quark-antiquark condensates and considering not only that  
these new chiral vacua are in turn affected by the presence of quark sources 
(the physics of chiral restoration/enhancement) \cite{nucmat} but also that
we might have excitations of these vacua along the "mexican hat" valley of a
continuum of chiral states connected with the "true" chiral vacuum, should
allow us to use the formalism of this paper to investigate and predict the
existence of signatures of these states in the nuclear environment. For this
we must have a model accurate enough both in the N-N sector and in the sector
of hadron spectroscopy. This paper constitutes a first step towards this goal.

The remainder of this paper is divided into four sections and three 
appendices. Section II is
devoted to the introduction of the specific Nambu Jonas-Lasinio effective
model that we will be using. The pion and Nucleon Salpeter equations will be
studied in section III. The overlap diagrams for the couplings $f_{\pi N N}$
are presented in section IV.  Finally in section VI we discuss the results. 
We have three appendices. In the first appendix some diagrams contributing to 
the $f_{\pi N N}$ are presented and discussed in detail. The color overlap 
results are given in the second appendix. In the last appendix  we discuss, 
both for the nucleon and the delta, the relevant flavor-spin overlaps.

\section{A SIMPLE CONFINING QUARK MODEL WITH CHIRAL 
SYMMETRY BREAKING}
Our quarks are Dirac fermions that interact with a simple effective 2-body
interaction \cite{AmeYao}
such as to simulate color confinement together with the introduction of 
a scale responsible for the actual particle sizes,
\begin{equation}\label{hamilton}
H = \int d^3 x \left( H_0({\bf x}) + H_I({\bf x}) \right)\;,
\end{equation}
\noindent
where $H_0$ is the Hamiltonian density of the Dirac field, and $H_I$ an
effective interaction term,
\begin{eqnarray}\label{hamilton1}
&&H_0({\bf x}) = \psi^{\dag}({\bf x}) \;(m_q\beta -i
               {\vec{\alpha} \cdot \nabla}   )
                 \;\psi({\bf x}) \;,\nonumber\\*
&&H_I({\bf x}) = \frac{1}{2} \int\!d^3y \:V_I ({\bf x -y}) \;\bar {\psi}
 ({\bf x})
    \frac{\lambda^a}{2} \gamma ^{0}\psi({\bf x}) \;\bar {\psi} ({\bf y}) 
    \gamma ^{0}\frac{\lambda^a}{2}
    \psi({\bf y}) \;,\nonumber\\*
&&V_I ({\bf x -y})=\frac{-3}{4}[K_{0}^3 ({\bf x-y})^2-U] \delta (t_{x}-t_{y}).
\end{eqnarray}
\noindent
The $\lambda^a$'s are the Gell-Mann color matrices. U is a arbitrarily large 
constant. In Ref. \cite{BicKre} we have shown that physical processes 
involving color singlets are independent of U whereas any colored objects get   
an infinite mass when $U\rightarrow \ \infty$.
\par
Different and more complex interactions have also been used in the 
literature. Within the Wilson loop context, results for the effective 
microscopic interaction among heavy quarks are available  \cite{Brambilla} 
but unfortunately these results cannot be extended to the light quark sector 
due precisely to the problem of consistency with chiral symmetry. 
Nevertheless it is customary to make the approximation of dividing the quark 
effective potential in two additive terms, one dominated by the coulomb force 
due to one gluon exchange which is responsible for the short distance 
interaction among quarks and one which plays an important role in the heavy 
quark spectroscopy, and another term- the confining term-which is responsible 
for the long distance force among quarks. In fact this approximation is also 
effectively used when extracting the heavy quark potentials from the Wilson 
loop. 
 
Although there is a great flexibility 
in the choice of the effective confining potential for the light quark mass 
sector, it is however not completely arbitrary. First it must 
not only to comply with the requirements of the Ward identities but also 
to provide finite results in the colorless sector while pushing the masses 
of the colored states to infinity.  This is illustrated when we consider
a more general Dirac structure like,
\begin{equation}\label{hgen}
V=V_I ({\bf x -y})\left( s\ 1 \otimes 1 + p\ \gamma_{5} \otimes \gamma_{5} 
+ v\ \gamma^{\mu} 
\otimes \gamma_{\mu} + a\ \gamma^{\mu}\gamma_{5} \otimes 
\gamma_{\mu} \gamma_{5} + 
 \frac{t}{2} \sigma ^{\mu \nu} \otimes \sigma_{\mu \nu} \right)
\end{equation}
Then the axial and vectorial Ward identities would have constrained
the parameters s, p, a, v and t to obey the following equations,\cite{VilLiu} 
\begin{equation}\label{WarCons}
s+p+6 t=0;\;
s-p=0.
\end{equation}
Equation (\ref{WarCons}) implies that the scalar (s), pseudo-scalar(p) and  
tensor(t) interactions do not contribute either to chiral symmetry 
breaking or to the structure of the ground state hadrons.
Finally we must also have,
\begin{equation}\label{conf2}
4 v -4 a=-2 v-2 a;\; a=0
\end{equation}
in order to have both the Goldstone pion and 
the other color singlets infrared independent.  
 As for the shape of the potential, the linear potential has also  
been extensively studied. However it seems to yield a much too large 
hadronic radius.

The potential of equation (\ref{hamilton},\ref{hamilton1}) has been used to 
study the 
charmonium spectroscopy with a potential strength of 
$K_0=290 MeV$. Although the theoretical results did not differ 
too much from the experimental ones it was shown in Ref. \cite{VilLiu} in 
the context of a more general potential, that we will still need a coulomb 
force if we want to get a correct value for the R parameter (which is related 
with the mass splittings of the $^3P_J$ triplet). And this happened 
despite the fact that, $individually$, the theoretically obtained masses were 
quite good when compared with the experimental ones. It turns out that the 
charmonium spectroscopy \cite{BicKre} obtained with the simpler potential of 
equation (\ref{hamilton1}) does not fare too badly either, when we consider 
individual masses, so that although no one disputes the fact that for short 
distances a coulomb force is needed, we decided (also for the sake of 
mathematical simplicity) to discard this force as a first step towards a 
quantitative description of the pion-nucleon coupling.
 
Therefore the model embodied in  equations (\ref{hamilton}, \ref{hamilton1}) 
not only seems to sufficiently meet all the above general requirements but 
also yields reasonable results besides allowing relatively simple calculations.
A covariant version of this model has been developed
in Ref. \cite{Horvat} and so far it has only been applied to the study of the 
pion and kaon. For the simple harmonic confining potential the authors 
obtained a substantial improvement on the value of $f_{\pi}$ although still 
far from a quantitative agreement. Coupled channels should constitute another 
sizeable correction not only to $f_{\pi }$ but also to the hadronic decay 
widths and masses. Even without covariance we have shown \cite{KN} that 
coupled channels provide a very large correction to $f_{\pi}$ therefore 
substantiating the hope that covariance plus coupled channels might bring the 
$f_{\pi}$ close to its experimental value. The present calculation 
constitutes a preliminary step in the microscopic and Lorentz invariant 
calculation of the $f_{N\pi N}$.In Eq.(~\ref{hamilton1}), the field operator 
$\psi ({\bf x})$ is defined as,
\begin{equation}
\psi ({\bf x})  = \int \frac{d^3p}{(2\pi)^{3/2}} \left[ u_s({\bf p})
   b ({\bf p}) +v_s({\bf p})d^{\dag} (-{\bf p})\right] e^{i{\bf p}
   \cdot{\bf x}} \;.
\end{equation}
 
\noindent
 $b$ and $d$ are respectively the quark and antiquark Fock space operators and
they carry indices for flavor, spin and color. Summation over repeated indices
is assumed. The spinors $u$ and $v$, together with the Fock space operators 
differ from those used in free Dirac theory and are given by, 
\begin{eqnarray}\label{spinors}
u_s({\bf p}) &=& \frac{1}{\sqrt{2}}\left[ [1+\sin \varphi (p)]^{\frac{1}{2}} 
+ [1-\sin \varphi (p)]^{\frac{1}{2}}\widehat{\bf p}\cdot
\vec{\alpha}\right] u_{s}^0 \;,\nonumber\\
v_s({\bf p}) &=& \frac{1}{\sqrt{2}}\left[ [1+\sin \varphi (p)]^{\frac{1}{2}}
- [1-\sin \varphi (p)]^{\frac{1}{2}}\widehat{\bf p}\cdot
\vec{\alpha}\right] v_{s}^0 
\end{eqnarray}
\noindent
In Eq.(\ref{spinors}) $u_s^0$ and $v_s^0$ are spinor eigenvectors of 
$\gamma_0$ corresponding to eigenvalues $\pm 1$. 
The function $\varphi(p)$ is called the {\em chiral angle} and indexes 
the different Fock spaces compatible with the Pauli Principle. 
This chiral angle has been studied in Ref.\cite{BicRib2,AmeYao} and is 
a solution of the {\em mass gap equation},
\begin{equation}\label{massgap}
(k^2 \varphi')'=2 k^3 \sin (\varphi )-\sin (2 \varphi )\;,
\end{equation}
where k is a dimensionless quantity in units of $K_0=1$. 
This unit will be used in the remainder of this paper.
In Fig.\ref{fchir} we plot $\sin (\varphi (k))$ as a function of k.

Once this mass gap equation is solved, the quark and
antiquark propagators can be diagonalized and it turns out that 
it is simpler to work in the spin representation rather than in the
Dirac representation. As it will be clear this representation 
will turn out to be the simplest representation if we want to study the 
$ \pi\ n$ or $\Delta$ couplings. In the spin representation we have for the 
quark energy E(k),
\begin{equation}\label{EK}
E(k) = k \cos (\varphi )-\frac{{\varphi'}^2}{2}-
\frac{{\cos^2 (\varphi )}}{k^2}
\end{equation}
The Feynman rules for the spin representation are given in Fig. \ref{ffeyn}.  
We have for the vertices,
\begin{eqnarray}\label{FeyVer}
&&{u^\dagger}_{s1}({\bf k1}) u_{s2}({\bf k2})=\frac{1}{2} \left\{
[\sqrt{1+S1} \; \sqrt{1+S2}+\sqrt{1-S1} \; \sqrt{1-S2} ({\hat {\bf k}}1 
\cdot
{\hat {\bf k}}2 )] \delta_{s1,s2}\right\}\nonumber\\*
&&\left. \hspace{25 mm} +\sqrt{1-S1}\; \sqrt{1-S2}\: (i {\vec \sigma} \cdot
{\hat {\bf k}}1 \times {\hat {\bf k}}2 )_{s1s2} \right\}\nonumber\\*
&&{v^\dagger}_{s1}({\bf k1}) v_{s2}({\bf k2})=\frac{1}{2} \left\{
[\sqrt{1+S1} \; \sqrt{1+S2}+\sqrt{1-S1} \; \sqrt{1-S2}\: ({\hat {\bf k}}1 
\cdot
{\hat {\bf k}}2 )] \delta_{s1,s2}\right\}\nonumber\\*
&&\left. \hspace{25 mm} +\sqrt{1-S1}\; \sqrt{1-S2}\: (i {\vec \sigma} \cdot
{\hat {\bf k}}1 \times {\hat {\bf k}}2 )^{*}_{s1s2} \right\}\nonumber\\*
&&u^{\dagger}_{s1}({\bf k1}) v_{s2}({\bf k2})=\frac{-1}{2} \left\{ 
[\sqrt{1-S1} \; \sqrt{1+S2}\: {\hat {\bf k}}1-
\sqrt{1+S1} \; \sqrt{1-S2}\: {\hat {\bf k}}2] \;
(i {\vec \sigma} \sigma_2)_{s1s2} \right\} \nonumber\\*
&&v^{\dagger}_{s1}({\bf k1}) u_{s2}({\bf k2})=\frac{1}{2} \left\{ \;
[\sqrt{1-S1} \; \sqrt{1+S2}\: {\hat {\bf k}}1-
\sqrt{1+S1} \; \sqrt{1-S2}\: {\hat {\bf k}}2] \;
(i {\vec \sigma} \sigma_2)^{+}_{s1s2}\right\}.\nonumber\\*
\end{eqnarray}
$S1$ and $S2$ stand respectively for $\sin (\varphi ( k1))$ and 
$\sin (\varphi (k2))$. As usual $\vec \sigma$ represents the Pauli 
matrices vector $\vec \sigma =\{ \sigma_1 ,\sigma_2 ,\sigma_3 \}$. The 
subscripts 
$s1,s2$ stand for the spin projections of the spinors and it is not hard to 
see that they can be put in a one-to-one relationship with the matrix 
elements of $ i\vec \sigma \sigma_2$ or, for that matter, with any other such 
vector.   Notice that the last two vertices which represent 
$q-\overline{q}$ 
pair creation or annihilation (the last vertex) are homogeneous functions of
$\vec \sigma $. This fact alone will entail a derivative coupling of the type
$\vec \sigma \cdot {\bf P}$ for the $N\pi ({\bf P}) N$ coupling.
For the propagators we have,
\begin{equation}\label{FeyPro}
{\cal S}_{q}({\bf k},w)={\cal S}_{\overline{q}}({\bf k},w)=
\frac{i}{w-E(k)+i \epsilon}
\end{equation}
Equipped with these rules we can now proceed to construct the
Salpeter amplitudes for arbitrary hadronic processes. Before concluding 
this section we would like to make a few remarks concerning some aspects of 
the physical picture embodied in the Hamiltonian (\ref{hamilton1}) and the 
Valatin-Bogoliubov transformed Dirac spinors of equation (\ref{spinors}). In 
many respects it is a picture similar to the BCS theory of superconductivity. 
For details on how to construct a BCS-like vacuum as a coherent superposition 
of $^3P_0$ quark-antiquark pairs see reference \cite{BicRib2}.  
Here as an illustrative example let us consider the quasi-quark creation 
operator $b^\dagger ({\bf p})$ with definite quantum numbers in spin, flavor 
and color. As in BCS, it represents in the usual Fock 
space (with a fermion empty vacuum) a superposition of a quark 
with those quantum numbers with a  coherent state made of ``Cooper-like'' 
$^3P_0$, color singlet, quark-antiquark pairs. And because 
of the Pauli principle, this coherent state is made of all the possible 
$^3P_0$ quark pairs with the single exception of that $^3P_0$ pair which 
would contain a quark with the same quantum numbers. In the same manner a
bound state (a meson or a baryon) will be a superposition of quarks (and/or 
antiquarks) with  $^3P_0$ coherent states built in such a way as to satisfy 
the Pauli exclusion principle. The usefulness of the Valatin-Bogoliubov 
transformation stems from the fact that it allows us to ``forget'' the details 
of the physical vacuum and therefore to treat complicated quark bound states 
(which are quite different from ordinary pure bound states) as normal bound 
sates of quasi-quarks with the information on the details of the physical 
vacuum stored inside the Dirac spinors in the form of a $chiral\ angle$. The 
Dirac structure of the quark spinors does not change this picture in any 
essential way provided we use Bethe-Salpeter equations to study these 
quasi-quarks bound states. For simplicity, the word quark will be used in the 
remainder of this paper to mean a quasi-quark.

\section{THE  PION, NUCLEON AND DELTA SALPETER EQUATIONS}
The hadronic Salpeter equations can be thought as the generating
equations for the Dyson series of the correspondent hadronic T-matrix
(see Ref. (\cite{BicRib3}). In Fig. \ref{fDyson} we depict this 
correspondence
for one single meson $q-\overline{q}$ bound state. The associated Salpeter
equation for an arbitrary meson $\Phi=(\Phi^{+},\Phi^{-}) $ with four
momentum $(E,{\bf P})$ is given by two coupled equations, one for the
positive-Energy component of the Salpeter amplitude $\Phi^{+} $
\begin{eqnarray}\label{PioSal+}
\Phi_{s1,s2} ^{+}({\bf k,\;P})=&&\int \frac{d^3 k'd w}{(2 \pi )^4}
{\cal S}_{q}({\bf k'}+\frac{{\bf P}}{2}\;,\frac{E}{2}+w)
{\cal S}_{\overline{q}}
({\bf -k'}+\frac{{\bf P}}{2}\;,\frac{E}{2}-w)
\; (-i V({\bf k-k'}))\;\times \nonumber\\*
&&\left\{
[u^{+}_{s1}({\bf k}+\frac{{\bf P}}{2}\;)
  u_{s3}({\bf k'}+\frac{{\bf P}}{2}\;)]
\ [-v^{+}_{s4}({\bf k'}-\frac{{\bf P}}{2}\;)
  v_{s2}({\bf k}-\frac{{\bf P}}{2}\;)]
  \Phi_{s3,s4} ^{+}({\bf k',\;P})\right. \nonumber\\*
&&\left. -[u^{+}_{s1}({\bf k}+\frac{{\bf P}}{2}\;)
  v_{s4}({\bf k'}+\frac{{\bf P}}{2}\;)]
\ [u^{+}_{s3}({\bf k'}-\frac{{\bf P}}{2}\;)
  v_{s2}({\bf k}-\frac{{\bf P}}{2}\;)]\Phi_{s3,s4} ^{-}({\bf k',\;-P})
\right\}\nonumber\\* 
\end{eqnarray}
and a similar equation for the negative energy component $\Phi^{-} $,
\begin{eqnarray}\label{PioSal-}
\Phi_{s1,s2} ^{-}({\bf k,\;P})=&&\int \frac{d^3 k'd w}{(2 \pi )^4}
{\cal S}_{q}({\bf k'}+\frac{{\bf P}}{2}\;,\frac{E}{2}+w)
{\cal S}_{\overline{q}}
({\bf -k'}+\frac{{\bf P}}{2}\;,\frac{E}{2}-w)
\; (-i V({\bf k-k'}))\;\times \nonumber\\*
&&\left\{
[u^{+}_{s3}({\bf k'}+\frac{{\bf P}}{2}\;)
  u_{s1}({\bf k}+\frac{{\bf P}}{2}\;)]
\ [-v^{+}_{s2}({\bf k}-\frac{{\bf P}}{2}\;)
v_{s4}({\bf k'}-\frac{{\bf P}}{2}\;)]
  \Phi_{s3,s4} ^{-}({\bf k',\;P})\right. \nonumber\\*
&&\left. -[v^{+}_{s4}({\bf k'}-\frac{{\bf P}}{2}\;)
  u_{s1}({\bf k}+\frac{{\bf P}}{2}\;)  ]
\ [v^{+}_{s2}({\bf k}-\frac{{\bf P}}{2}\;)
  u_{s3}({\bf k'}-\frac{{\bf P}}{2}\;)]\Phi_{s3,s4} ^{+}({\bf k',\;-P})
\right\}\nonumber\\* 
\end{eqnarray}
In what follows we denote by $\Phi ({\bf P})$ 
the energy-spin doublet $(\Phi^{+}(k,{\bf P}),\Phi^{-}(k,{\bf -P}))$,
therefore omitting the $q-\overline{q}$ internal momentum k and the spins.
We will also denote the Taylor series in ${\bf P}$ of a given function
$F({\bf k},\;{\bf P})$ by $\sum F_n$ instead of the usual
$\frac{1}{n!}\sum {\bf f}_n\cdot [{\bf P}]^{n}$, with $[{\bf P}]^{n}$ being
a shorthand notation for a tensor in $P$ of degree n.

Notice that in Eqs. (\ref{PioSal+},\ref{PioSal-}) we can integrate out
the quark and antiquark propagators,
\begin{equation}
\int \frac{d w}{2 \pi}
{\cal S}_{q}({\bf k},\frac{\pm E}{2}+w) {\cal S}_{\overline{q}}
({\bf -k},\frac{\pm E}{2}-w)=\frac{i}{\pm E -E_{q} ({\bf k})-
E_{\overline{q}} ({\bf k})}
\end{equation}
therefore allowing us to rewrite Eqs. (\ref{PioSal+}) and  (\ref{PioSal-}) as,
\begin{equation}\label{PioSal1}
\left(H({\bf k},\; {\bf P})\ -\ E_{meson}({\bf P})\; I
\right)
\left[ \begin{array}{c} \Phi^{+}({\bf k},\; {\bf P})\\
\Phi^{-}({\bf k},\; {\bf -P})
\end{array}\right]\;=\;0\ ;\ I\ =\ 
\left[
\begin{array}{cc} 1&0\\
0&-1
\end{array}\right],
\end{equation}
were $H({\bf k},\; {\bf P})$ is a two by two matrix,
\begin{equation}\label{aha}
H({\bf k},\; {\bf P}) =
\left[ \begin{array}{cc}
H^{++}({\bf k},\; {\bf P})&H^{+-}({\bf k},\; {\bf P})\\
H^{-+}({\bf k},\; {\bf P})&H^{--}({\bf k},\; {\bf P})\end{array} \right],  
\end{equation}
which depends in the center-of-mass momentum P. The bare masses of mesons
are just the eigenvalues $E_0=M$ of $H(0)=H_{0}$. 
We have for $H^{(+;-)\; (+;-)}({\bf k},\; {\bf P})$,
\begin{eqnarray}\label{H++H+-}
&H^{++}=H^{--}&=
E(p1)+\frac{1}{4}\;\left[  \left( \varphi' (p1)^2+\frac{4(1-S1)}{p1^2}\right)
+\frac{1}{6}\; \frac{(1-S1)\; (1-S2)}{p1^2 p2^2} {\bf p1} \cdot {\bf p2}  
\right] (\vec \sigma_{qk1}\cdot \vec \sigma_{qk2}) +\nonumber\\*
&&{\bf p1} \rightarrow {\bf p2} + h.r.t. \nonumber\\*
&H^{+-}=H^{-+}&=
-\frac{1}{12}\; \left[ \left( \varphi' (p1)^2+\frac{C1}{p1}\right)\; 
\left( \varphi' (p2)^2+\frac{C2}{p2}\right) 
\frac{({\bf p1} \cdot {\bf p2})^2}{p1^2 p2^2}-\right. \nonumber\\*
&&\left. \frac{1}{3} \left( \varphi' (p1)+\frac{C1}{p1}\right) \; 
\frac{C2}{p2}+
\frac{C1 C2}{2 p1 p2}\right](\vec \sigma_{qk1}\cdot \vec \sigma_{qk2}) 
+{\bf p1} \rightarrow {\bf p2}+h.r.t.
\end{eqnarray}
with ${\bf p1}={\bf k}+{\bf P}/2$ and ${\bf p2}={\bf k}-{\bf P}/2$ being 
respectively the momenta of the quark 
and the antiquark and ${\bf P}$ the momentum of the pion. $h.\; r.\; t.$ 
stands for 
higher rank tensors which will not contribute for the pion wave function and 
$S1$ represents $\sin (\varphi (p1))$, $C2$ represents $\cos (\varphi (p2))$ 
and so on. $\vec \sigma_{qk1}$ and $\vec \sigma_{qk2}$ stand for two Pauli 
matrices vectors acting respectively in quark 1 and antiquark 2.
In the case of the Goldstone pion we have for $[H_{0}-E_{0}] |\phi \rangle$,
\begin{equation}\label{PioHam}
\left\{ \left( -\frac{d^2}{dk^2}+2 E(k)\right)
\left[ \begin{array}{cc}1&0\\0&1\end{array}\right]
+\left(\frac{\varphi'^2}{2}+\frac{\cos (\varphi )}{k^2}\right)\;
\left[\begin{array}{cc} 1&1\\1&1\end{array}\right]-M\;
\left[\begin{array}{cc} 1&0\\0&-1\end{array}\right]\right\}\;
\left[\begin{array}{c} \nu^{+} (k)\\\nu^{-} (k)\end{array}\right]=0
\end{equation}
where $\nu ^{\pm}=k\; \Phi^{\pm}$.
In the limit of vanishing current quark masses we have that
$\Phi_{0}^\dagger I\Phi_{0} =0$. In this limit we also have that
$M=0$ and that $\Phi_{0}^\dagger H_1\Phi_{0} =0$.

Therefore in order to find the energy and the norm of the Goldstone
pion to order P, it turns out to be necessary to expand equation 
(\ref{PioSal1}) to second order in the pion momentum $\vec P$.  We obtain, 
\begin{eqnarray}\label{EqSalExp}
&&\left( H_0-E_0 I\right) \Phi_0=0 \nonumber \\       
&&\left( H_0-E_0 I\right) \Phi_1
+\left( H_1- E_1 I\right) \Phi_0=0 \nonumber \\
&&\left( H_0-E_0 I\right) \Phi_2
+2 \left( H_1- E_1 I\right) \Phi_1   
+\left( H_2-E_2 I\right) \Phi_0=0
\end{eqnarray}                                                       
Then we can obtain both the pion energy $E_{1}$, correct up to first order 
in P, 
\begin{equation}\label{EnerPi}
\Phi_{0}^\dagger \left( H_1- E_1 I\right) \Phi_1+  
\Phi_{0}^\dagger\; H_2\; \Phi_0=0
\end{equation}
and the pion Salpeter wave function normalization which is given by,
\begin{equation}\label{NormPi}
{\cal N}^2 = \Phi_{1}^\dagger I\Phi_{0}  + \Phi_{0}^\dagger I\Phi_{1}
\end{equation}
To obtain the desired pion Salpeter amplitude, we need to apply 
the above general formalism for our particular model. Using the Feynman 
rules defined in section II, we are able to obtain the matrix elements of 
Eq (\ref{aha}).
\begin{eqnarray}\label{H0H1H2}
H_0^{++}=&&H_0^{--}=2E-\bigl\{ \Delta 
-{1 \over 2} (\varphi'^2+4{1-S \over k^2})
-{1 \over 3} {(1-S)^2 \over k^2} \vec \sigma_{qk1} \cdot \vec \sigma _{qk2}^t
+  h.\; r.\; t.\bigr\} \nonumber \\
H_0^{+-}=&&H_0^{-+}= {1 \over 3} ({\varphi'^2 \over 2}+{C^2 \over k^2})
 \vec \sigma_{qk1} \cdot \vec \sigma _{qk2}^t
+ h.\; r.\; t.\\
H_1^{++}=&&H_1^{--}= -\bigl\{  
{1 \over 4} {(1-S)^2 \over k^3} 
{\bf P} \cdot \hat {\bf k} \times (\vec \sigma_{qk1} \times \vec 
\sigma _{qk2}^t)
\nonumber \\ &&
+{1 \over 2}({C \varphi' \over  k}+2{1-S \over k^2} )
{\bf P} \cdot \hat {\bf k}\: [i \hat {\bf k} \cdot (\vec \sigma_{qk1} - 
\vec \sigma _{qk2}^t) ]
-{1-S \over {2\; k^2}} 
{\bf P} \cdot  i (\vec \sigma_{qk1} - \vec \sigma _{qk2}^t)  
\times  \nabla_{\bf  k}
\bigr\} \nonumber \\
H_1^{+-}=&&H_1^{-+}= 
{1 \over 4 k} ( \varphi' + {C \over k} )^2
{\bf P} \cdot \hat {\bf k} \times (\vec \sigma_{qk1} \times \vec 
\sigma _{qk2}^t)
\nonumber \\
H_2^{++}=&&H_2^{--}=
\sum \frac{d^2 H^{++}}{(d\;P_i)^2} P_i^2\; (i=1,2,3)\nonumber \\
H_2^{+-}=&&H_2^{-+}=
\sum \frac{d^2 H^{+-}}{(d\;P_i)^2} P_i^2\; (i=1,2,3)
\end{eqnarray}
As usual $S$ stands for $\sin (\varphi ({\bf k}))$ and $C$ for 
$\cos (\varphi ({\bf k}))$. $k$ represents the $q-\overline{q}$ relative 
momentum and ${\bf P}$ the momentum of the pion. Finally we solve the above 
equations to obtain the boosted pion Salpeter wave function, correct up 
to first order in P,
\begin{eqnarray}\label{pi+pi-}
&&\Phi^+={\cal N}^{-1}\left(S + E_1 f_1 + i
g_1 {\bf P} \cdot \hat {\bf k} \times \vec \sigma \right) \;\chi_{\pi }\; 
{\cal S}_{color}
\nonumber \\
&&\Phi^-={\cal N}^{-1}\left(-S + E_1 f_1 - i
g_1 {\bf P} \cdot \hat {\bf k} \times \vec \sigma \right) \;\chi_{\pi }\;
{\cal S}_{color}
\nonumber\\* 
&&{\cal N}^2=4E_1\;\int \frac{d^3 k}{(2 \pi)^3} S\;f_1=E_1 a^2\nonumber\\*
&&E_1=\frac{2P}{a}\;\sqrt{\int \frac{d^3k}{(2 \pi )^3}\; S^2 \sum
\left( \frac{d^2 H^{++}}{(d\;P_i)^2}- \frac{d^2 H^{+-}}{(d\;P_i)^2}\right)}
(i=1,2,3)
\end{eqnarray}

In Eq. (\ref{pi+pi-}) $\chi_{\pi }$ and ${\cal S}$ describe respectively the 
spin-flavor pion wave function given in Eq. (\ref{spin-fla}), and the usual 
mesonic color singlet wave function.
It is important to notice that the pion normalization goes as $\sqrt{(E_1)}$. 
This fact will be crucial to retrieve the well known OPEP potential in the 
static approximation-see Eq. (\ref{OPEP}). The constant $a$ equals $0.16\;  
K_0$.

It is convenient to
introduce the notation $\Phi ^{\pm}={\cal N}^{-1}(\phi _{0}^{\pm}+
\phi _{1}^{\pm})$
to be able to separate the contributions of the pion at rest and its
associated boost correction to the $N\pi N$ coupling. From Eq.
(\ref{pi+pi-}) we see that in the rest frame the pion  Salpeter space-wave
function $\phi _{0}$, is simply given by the $\sin (\varphi ) $
whereas
$\phi_1 ={\cal N}\left(E_1 f_1 + i g_1
{\bf P} \cdot \hat {\bf k} \times \vec \Sigma \right)$. $f_1$ and $g_1$ are the
solutions of the differential equations,  
\FL
\begin{eqnarray}\label{f1g1eq}
&&\left[ -\frac{d^2}{dk^2}+2 k\; C \right] f_1=
k\;  S\nonumber\\*
&&\left[ -\frac{d^2}{dk^2}+2 k\;C+\frac{2 S^2}{k^2}\right] g_1= 
\hspace{2mm}\frac{i}{2 k^3} \left[ 2 k C\varphi ' + S
(2 S^2-2S-k^2 {\varphi '}^2- 4 k C \varphi ')\right].
\end{eqnarray}
\par
The solutions $f_1$ and $g_1$ of Eq.(\ref{f1g1eq}) are given in 
Fig. \ref{f1g1}. This result is the same as that obtained in 
ref \cite{AmeYao}, provided we perform the following change,
\begin{equation}
g_1=-i (f_2-\frac{\sin (\varphi (k))}{2})
\end{equation}
\par
Surprisingly the baryon Salpeter equation turns out to be simpler than 
the corresponding pion Salpeter equation. This happens because in the 
baryon case, or indeed any other baryon, the associated Dyson series for the  
corresponding S-matrix with an instantaneous interaction, does not have 
negative energy channels (which would correspond to an $Bar-\bar Bar$
component
of the baryon propagator and therefore of negligible importance for the 
nucleon ground state in sharp contrast with the pion case) nor it has 
couplings to negative energy quark-positive energy diquark or positive 
energy quark-negative energy diquark channels due to color confinement. 
We also make the approximation that the baryon Salpeter wave function does
not change for small boosts due to its heavy mass. In Fig. \ref{fNucl} we 
show the N associated Salpeter equation. 
In what follows we will use N to represent either the nucleon or the delta. 
When need arises to distinguish the nucleon from the delta we will reserve n 
to denote the nucleon. The N Salpeter equation is then,
\FL
\begin{eqnarray}\label{bbary}
\lefteqn {[M\!-\!3E(p1)]\;\chi_{s_1s_2s_5}\; \psi({\bf p}1,
{\bf p}2,{\bf p}3)=} \\*
& &-2\int\! d^3q\; V({\bf q})\;[u_{s_1}^{\dag} ({\bf p}1)
u_{s_3}({\bf p}1\!-\!{\bf q}) u_{s_2}^{\dag}({\bf p}2)
u_{s_4}({\bf p}2\!+\!{\bf q})] \chi_{s_3s_4s_5} \psi({\bf p}1\!-\!{\bf q},
{\bf p}2\!+\!{\bf q},{\bf p}3) \;,\nonumber
\end{eqnarray}
with ${\bf p}1\!+\!{\bf p}2\!+\!{\bf p}3=0$. 
This equation can be solved variationally,
\begin{equation}
\delta_{\alpha} [\langle N_{\alpha}\;|M-{\cal H} |\;N_{\alpha} \rangle ] =0
\end{equation}
where ${\cal H}$ is a shorthand notation representing all the terms of 
Eq.
(\ref{bbary}) except for the eigenvalue M, $ \chi_{s_1s_2s_5}$ is given in 
Eqs.(\ref{wavespin},\ref{DELTA}) in appendix C.

For the momentum representation of the nucleon wave function
$\langle {\bf p}1, {\bf p}2,{\bf p}3| N_{\alpha}\rangle $ we used,
\begin{eqnarray}\label{Nsal}
&&\langle {\bf p}1, {\bf p}2,{\bf p}3| N\rangle=
\frac{e^{\displaystyle 
- \frac{{\bf \rho}^2 + {\bf \lambda}^2}{2 \alpha^2}}}{{\cal N}}\nonumber\\*
&&{\bf \rho} =\frac{{\bf p}1 -{\bf p}2}{\sqrt{2}},\;
{\bf \lambda} =\frac{{\bf p}1 +{\bf p}2- 2 {\bf p}3}{\sqrt{6}}
\end{eqnarray}
with $\rho $ and $\lambda $ the appropriate Jacobian variables. $\alpha $,
the inverse baryon radius, turns out to be the same for the nucleon and
the delta.
The total $n$, $\Delta$ wave function $\Psi_{N}$ is then given by,
\begin{equation}\label{totnucw}
\Psi_{N}=\langle {\bf p}1, {\bf p}2,{\bf p}3| N_{\alpha}\rangle \;
\langle flavor|N \rangle \; \langle color|N \rangle
\end{equation}
where $\langle flavor|N \rangle $ stands for the appropriated spin-flavor 
content of either the nucleon or the Delta ( see Eqs. (\ref{wavespin}-
\ref{DELTA}) and $\langle color|N \rangle $ for the usual normalized color 
Slater determinant. 
The spin representation is then the natural  representation for the 
reduction of the Salpeter equation to the Schrodinger-like equation 
(\ref{bbary}). 

The case of the linear confining potential has also been studied in
Ref. \cite{KN} and for completeness we simply quote the final
conclusions: We obtained  for the same nucleon bare mass a larger
nucleon core in the case of the linear confining potential than in the 
harmonic case. This result was then shown to be difficult to  
accommodate with the k-N exotic scattering which seems to favor a small 
nucleon core which is precisely what happened with the harmonic force.

\section{THE OVERLAP DIAGRAM FOR THE PION COUPLING}
When studying meson-baryon scattering we need to consider non exotic 
scattering amplitudes which are induced by quark-antiquark annihilation 
(or creation) amplitudes. The Feynman diagrams contributing to the
$n\pi n$ and $\Delta \pi n$ vertices are given in Fig. \ref{foverl}.
We need to evaluate,
\begin{eqnarray}\label{N-Pi-N}
&&f_{N\pi N}=\frac{m_{\pi }}{a}\;\vec v_{N\pi N} \cdot \vec {\cal T}
\end{eqnarray}                      
where $\vec v_{N\pi N}$ and $\vec {\cal T}$ are defined in appendices C 
( Eq. \ref{spinflave})and A (Eq. \ref{sfamputated1}, \ref{sfamputated2}) 
respectively. The constant $a$ pertaining to the pion Salpeter 
normalization was introduced in Eq. (\ref{pi+pi-}). It is
important to notice that the remainder of the pion normalization ${\cal N}$, 
$\sqrt{E_1}$ is absorbed in the pion propagator,
\begin{equation}\label{mipion}
{\cal P}=\frac{i}{w-E_{\pi}}
\end{equation}
We have that in this approximation $E_{\pi}=E_1$ and because in OPEP we have 
two $N\pi N$ vertices we get an overall energy dependence $1/E^2_1$ which is 
in accordance with the OPEP potential in the static approximation,
\begin{equation}\label{OPEP}
V^p_3=-\; (\frac{f_{N\pi N}}{m_{\pi}})^2 ( P^2-m_{\pi}^2)^{-1}
\; \vec \tau_1 \cdot \vec \tau_2 \; (\vec \sigma_{N1} \cdot {\bf P}) 
(\vec \sigma_{N2} \cdot {\bf P})
\end{equation}
where $\vec \sigma_{N1}$ represents the vector of Pauli matrices acting in 
nucleon N1 as a whole. The same applies to $\sigma_{N2}$.

$(\vec v_{n\pi n} \cdot \vec {\cal T})\; (\vec \sigma_{n} \cdot {\bf P})\; 
\tau_{n}$ and 
$(\vec v_{n\pi \Delta} \cdot \vec {\cal T})\; (\vec S \cdot {\bf P})\; 
\vec T$ are 
examples of the overlap ${\cal O}=\langle N |H_I|(\phi _0 +\phi _1) N\rangle$, 
where $H_I$ is the Hamiltonian of Eq.(\ref{hamilton}).
The only contribution to ${\cal O}$ is provided by the
term of $H_I$ with a single $q-\overline{q}$ annihilation-see Fig.
\ref{foverl}.
We can expand ${\cal O}$ up to first order in the pion momentum to
obtain,
\begin{eqnarray}\label{Ogene}
&&{\cal O}=\langle N |h_0+h_1 |(\phi _0 +\phi _1) N\rangle \nonumber\\*
&& {\cal O}={\cal O}_a + {\cal O}_b\nonumber\\*
&&{\cal O}_a=\langle N |h_1 |\pi _0  N\rangle  ,\;
{\cal O}_b=\langle N |h_0|\pi _1 N\rangle
\end{eqnarray}
where h stands for $H_I$. Notice that $\langle N |h_0 |\phi _0 N\rangle =0$.
That is, at rest the pion decouples from the nucleons. The reason for this
is quite simple. From Eq. (\ref{FeyVer}) the $q-\overline{q}$ annihilation
vertex is of the form $\vec \sigma \cdot \vec k$ being $\vec k$ one of the 
momenta flowing in either the quark or the antiquark leg. In turn these 
momenta are sums of internal loop
momenta ${\bf k}^i=\{ {\bf k},{\bf k'} ,{\bf k}''\}$ and the external momenta 
which in this case turns out to be {\bf P} the pion momentum. 
Then upon integration in the internal momenta loops all the terms which are
of the type $\vec \sigma \cdot {\bf k}^i$ will disappear while the terms 
homogeneous in $\vec \sigma \cdot  {\bf P} $ are the only surviving terms.
In the appendices we derive ${\cal O}_a $ both for 
$n\pi n$ and $n\pi \Delta$. Due to the smallness of $g_1$, ${\cal O}_b$ turns 
out to be too small and it will be omitted henceforth. In appendix A it is 
shown that in the case of $n\pi \Delta$ we have an extra contribution for 
${\cal O}_a$ which is absent in the case of $n\pi n$. We will denote it by 
${\cal O}'$. It will also turn out to be quite small.

We obtained the following values for the adimensional quantities 
${\cal O}_{fs}$ and ${\cal O}'_{fs}$,
\begin{equation}\label{oo}
{\cal O}_{fs}=0.54;\; {\cal O}'_{fs}\simeq 0
\end{equation}
where ${\cal O}_{fs}$ and ${\cal O}'_{fs}$ stand for ${\cal O}$ and 
${\cal O}'$ amplitudes with the spin-flavor terms amputated-see Eqs. 
(\ref{sfamputated1}, \ref{sfamputated2}). 
\section{COMPARISON WITH EXPERIMENT AND DISCUSSION}
The experimental values for $f_{n\pi n}$ and $f_{n\pi \Delta}$ are 
respectively,
\begin{equation}
f_{n\pi n}= 1.0;\; f_{n\pi \Delta}\simeq 2.1
\end{equation}

We can use the results of Eq. (\ref{N-Pi-N}) to write the following 
set of equations,
\begin{eqnarray}
&& \frac{f_{n\pi n}}{m_{\pi}}=\frac{5}{9}\;\frac{{\cal O}_{fs}}{a}\nonumber\\*
&& \frac{f_{n\pi \Delta}}{m_{\pi}}=\frac{2\sqrt{2}}{3}\frac{{\cal O}_{fs}}{a} +
\sqrt{2}\frac{{\cal O}'_{fs}}{a}
\end{eqnarray}

Using the values of Eq. (\ref{oo}) for ${\cal O}$, ${\cal O}'$ and $a$ we are 
able to get for $f_{n\pi n}$ and $f_{n\pi \Delta}$,
\begin{equation}
f_{n\pi n}=1.9\; \frac{m_{\pi }}{K_0};\;f_{n\pi \Delta}=3.2\; 
\frac{m_{\pi }}{K_0}
\end{equation}

If we use the value of $K_0=247 MeV $ of ref. \cite {AmeYao} we obtain 
the theoretical results,
\begin{equation}
f_{n\pi n}=1.0;\;f_{n\pi \Delta}=1.8 
\end{equation}

These results are surprisingly good. 

We have seen that in order to explain the K-n exotic phase shifts we 
needed a smaller bare nucleon and therefore a larger $K_0\simeq 400 MeV$. With 
 this $K_0$ we would get  $60\%$ of the values just obtained. 
However we still have to consider the effects of Lorentz covariance and the 
contribution of the pion cloud around the bare nucleon. The study of these 
two effects will constitute a necessary step in this program of obtaining a 
quantitative microscopic description of low-energy hadronic phenomena. 
Nevertheless it is 
already remarkable that such a simple model (with only one scale $K_0$  apart 
from the quark masses) should yield results (ranging from hadronic 
spectroscopy to the coupling of pions to nucleons) which are not obviously 
wrong. This is more so if we take in consideration that this model is able to 
unify in the same description, essentially depending in the chiral  
angle $\varphi $, the (exotic) hadronic repulsion like for 
instance the nucleon-nucleon repulsive core (which is of the same nature than 
the k-n exotic repulsion) and the  n-n peripheral attraction 
mediated by pions. Retrospectively it is not hard to understand 
why the ``naive'' $^3P_0$  model (QPCM)\cite{QPCMM} for strong decays should 
perform so well. It is the minimal model which contains overlap kernels and 
satisfies parity conservation. In this sense any microscopic model (like the 
present one) which produces a pion derivative coupling can be simulated by 
QPCM. The pion momenta ${\bf P}$ which is present in the hadron-pion coupling 
$\vec \sigma \cdot {\bf P}$ and which forces a P-wave pion-hadron outgoing 
relative wave function can be made to have originated in an ``incoming'' 
cluster P wave $q-\bar q$ bound state. Parity conservation will force it to 
be a $^3P_0$ $q-\bar q$ bound state. Its amplitude of occurrence, in the 
literature denoted by $\gamma$, can afterward be fitted to data. This is 
clearly seen if we use the graphical rules \cite{GraphicalRules,KN} to 
evaluate these overlap kernels. Of course with such a minimal model one 
looses any connection with spectroscopy (hadron bare masses) and the physics 
of chiral symmetry breaking $S \chi  SB$.   

The $N-\bar N$ scattering  constitutes another area where 
the present model could and should be tested. At this stage we can already  
anticipate that the present model will again produce results which can be 
simulated by QPCM. To see this it is sufficient to notice 
that from the point of view of overlap kernels and as a qualitative guide 
we can lump together either the two spectator $\bar q$ or the two 
spectator quarks as 
an effective extended quark or antiquark respectively and therefore 
understand the $N-\bar N$ scattering as a modified $N-\pi$ scattering. The 
present calculation will therefore constitute a prerequisite to the  
calculation of the more complicated $N-\bar N$ scattering. The spectroscopy 
and scattering reactions for higher angular momenta will 
constitute another class of stringent tests notably in what concerns 
the old problem of van der Waals forces which we feel can only be 
realistically compared with experiment in the framework of covariant 
improvement of this model (retardation ).  
\appendix\section{ The evaluation of ${\cal O}$. An example}
First a note on notation. We denote by ${{(N\;\pi )}_{ij}}^{+}$ a
$N-\pi -N$ diagram for positive pion E-energy with a potential
insertion between quarks i and j. For the negative E-energy pion,
$N\pi N$ will be denoted by ${{(N\;\pi )}_{ij}}^{-}$.
In this appendix we will evaluate in detail the diagrams
${{(N\;\pi )}_{11}}^{+}$ and ${{(N\;\pi )}_{11}}^{-}$. The other
diagrams can be evaluated in similar fashion and its
derivation will be omitted. The Fourier transform of the potential 
$ K_{0}^3\; r^2$ is given 
by,
\begin{equation}\label{poten}
V({\bf k})=- (2 \pi)^3 K_{0}^3 \Delta_{{\bf k}} \delta ({\bf k})
\end{equation}
 The quark momenta we used are the following,
\begin{eqnarray}\label{momcov}
&Incoming Nuc.\ &{\bf p}1  ={\bf k}'+{\bf k}''-\frac{{\bf P}}{2};\ 
{\bf p}2  =-{\bf k}'+{\bf k}''-\frac{{\bf P}}{2};\ 
{\bf p}3  =-2\;{\bf k}'';\nonumber\\* 
&Outgoing Nuc.\ &{\bf q}1  ={\bf k}'+{\bf k}''+\frac{{\bf P}}{2};\ 
{\bf q}2  =-{\bf k}'+{\bf k''}+\frac{{\bf P}}{2};
\ {\bf q}3  =-2 \;{\bf k}''\nonumber\\* 
&Pion&{\bf p}4=-{\bf k}-{\bf k}''+\frac{{\bf P}}{2};\ 
{\bf q}4={\bf k}+{\bf k}''+\frac{{\bf P}}{2}
\end{eqnarray}
\par
We also discard the terms depending on $g_1$ because they are 
negligible and therefore will not affect the final result.
\par 
Integrating by parts, we can get rid of the $\delta$ in the potential
(\ref{poten}) and we have only to consider the effect of the laplacian and 
gradients on the vertices. We have the following cases to consider,
\begin{eqnarray}
{\cal O}_{11}^{+}=&&{\langle \lambda \cdot \lambda \rangle }_{11}
\int \frac{dk^3 dk'^3 dk''^3}{(2 \pi )^9} \left[  
[v^{\dagger}_{s3}({\bf p}4) \Delta_{{\bf k'}}\;(u_{s1}({\bf p}1))
+2 v^{\dagger}_{s3}( {\bf p}4) \; \nabla_{\bf k'}
(u_{s1}( {\bf p}1))\cdot  \nabla_{\bf k'}]\;
\delta_{s1''s1'}\right. \nonumber\\*
&&+\left.
2 [v^{\dagger}_{s3}({\bf p}4) \nabla_{\bf k'}(u_{s1}({\bf p}1))]
\cdot [  \nabla_{\bf k'}(u^{\dagger}_{s1''}( {\bf q}1))
u_{s1'}( {\bf q}4)] \right] \times \delta ^3({\bf k}-{\bf k'})
\nonumber\\*
&&[N({\bf q}1, {\bf q}2, {\bf q}3)^{\dagger } \;
\Phi \; N({\bf p}1, {\bf p}2, {\bf p}3)]
\end{eqnarray}
with,
\begin{eqnarray}\label{vertNPiN}
&&v^{\dagger}_{s3}( {\bf p}4) \Delta_{{\bf k'}}\;[u_{s1}
( {\bf p}1)]|_{k=k'}\;= 
\frac{1}{2} [\varphi''(p1)+\frac{2 \varphi'(p1)}{p1}+
\frac{2 \cos (\varphi(p1))} {p1^2}]\; \vec 
\Sigma_{s3s1}^{*} \cdot\hat {\bf p}1\nonumber
\\*
&&v^{\dagger}_{s3}({\bf p}4) \; \nabla_{\bf k'}[u_{s1}({\bf p}1)]|_{k=k'}\;= 
\frac{1}{2}[\varphi' (p1)+\frac{\cos (\varphi (p1))}{p1}]
(\vec \Sigma_{s3s1}^{*} \cdot \hat {\bf p}1) \hat {\bf p}1-
\frac{\cos (\varphi (p1))}{p1} \vec \Sigma_{s3s1}^{*}\nonumber\\*
&& \nabla_{\bf k'}[u^{\dagger}_{s1''}
( {\bf q}1)] u_{s1'}( {\bf q}4)|_{k=k'}\;= 
-\frac{i}{2} \frac{1-\sin (\varphi (q1))}{q1} 
(\vec \sigma_{s1''s1'}\times
\hat {\bf q}1).
\end{eqnarray}
$N({\bf p}1, {\bf p}2, {\bf p}3)$ is given by equations (\ref{Nsal}, 
\ref{totnucw}) and $\phi $ stands for the pion wave function of 
Eq. (\ref{pi+pi-}). 

The overlap ${\cal O}_{11}^{-}$ is given by,
\begin{eqnarray}
{\cal O}_{11}^{-}=&&{\langle \lambda \cdot \lambda \rangle }_{11}
\int \frac{dk^3 dk'^3 dk''^3}{(2 \pi )^9} \left[  
[u^{\dagger}_{s3}({\bf p}4) \Delta_{{\bf k'}}\;(u_{s1}({\bf p}1))
+2 u^{\dagger}_{s3}( {\bf p}4) \; \nabla_{\bf k'}
(u_{s1}( {\bf p}1))\cdot  \nabla_{\bf k'}]\;
\delta_{s1''s1'}\right. \nonumber\\*
&&+\left.
2 [u^{\dagger}_{s3}({\bf p}4) \nabla_{\bf k'}(u_{s1}({\bf p}1))]
\cdot [  \nabla_{\bf k'}(u^{\dagger}_{s1''}( {\bf q}1))
v_{s1'}( {\bf q}4)] \right] \times \delta ^3({\bf k}-{\bf k'})
\nonumber\\*
&&[N({\bf q}1, {\bf q}2, {\bf q}3)^{\dagger } \;
\Phi \; N({\bf p}1, {\bf p}2, {\bf p}3)]
\end{eqnarray}

For the laplacian it is not hard to see that the contributions of 
${(N\pi )_{11}}^{+}\; +\; {(N\pi )_{12}}^{+}\; +\; {(N\pi )_{13}}^{+}$ 
add to zero. The same happens for the negative energy pion amplitude. 
This is due to the fact that,
\begin{equation}\label{corsat}
\lambda \cdot (\lambda_1+\lambda_2+\lambda_3) f(q)=0 
\end{equation}
for any color singlet nucleon $N({\bf p1},{\bf p2},{\bf p3})$. We also have 
with all 
generality that the contributions of ${(N\pi )_{12}}^{[+,\; -]}$ are 
identical to 
the contributions of ${(N\pi )_{13}}^{[+,\; -]}$. This is a consequence of the 
antisymmetric properties of the incoming and outcoming nucleon wave functions.

Using the results of appendix B and after summing in the color degree of 
freedom we have,    
\begin{eqnarray}\label{casos}
&& {\bf Term\;a})\ [v^{\dagger}_{s3}({\bf p}1) \nabla (u_{s1}({\bf p}1))]
\cdot  \nabla (N \phi^{+} \;N)\nonumber\\*
&&\nonumber\\*
&& terms [+]: (N\;\pi )_{11}^{+}+(N\;\pi )_{12}^{+}+(N\;\pi )_{13}^{+}
\nonumber\\*
&&(-\frac{1}{\sqrt{3}}) \; \frac{1}{2} [\varphi' (p1)+
\frac{ \cos (\varphi (p1)) } {p1} ]
(\Sigma_{s3s1}^{*} \cdot \hat {\bf p}1) \hat {\bf p}1-
\frac{\cos (\varphi (p1))}{p1} \vec \Sigma_{s3s1}^{*} \cdot
[ \nabla (N_{out})\;\Phi^{+} N_{in}]
\nonumber\\*
&&\nonumber\\*
&&terms [-]: (N\;\pi )_{11}^{-}+(N\;\pi )_{12}^{-}+(N\;\pi )_{13}^{-}
\nonumber\\*
&&(-\frac{1}{\sqrt{3}})\;\frac{1}{2}[\varphi' ( q1)+
\frac{\cos (\varphi (q1))}{q1}]
(\vec \Sigma_{s1's1''} \cdot \hat {\bf q}1) \hat {\bf q}1-
\frac{\cos (\varphi (q1))}{q1} \vec \Sigma_{s1's1''} \cdot
[N_{out}\;\Phi^{-} \nabla N_{in}]\nonumber\\*
&&\nonumber\\*
&& {\bf Term\;b})
\ 2 \left[v^{\dagger}_{s3}({\bf p}1) \nabla_{\bf k'}
(u_{s1}({\bf p'}))]
\cdot [ \nabla_{\bf k'}(u_{s1''}( {\bf p'}))
u_{s1'}( {\bf p}1)] \right]  
\nonumber\\*
&&\nonumber\\*
&& terms [+]: (N\;\pi )_{11}^{+} 
\nonumber\\*
&&2 [-\frac{i}{2} \frac{1-\sin (\varphi( q1))}{q1}] 
\left[ \frac{1}{2}
[\varphi' (p1)+ \frac{\cos (\varphi (p1))}{p1} (
\vec \Sigma_{s3s1}^{*} \times \vec \sigma_{s1''s1'} )\cdot [\frac{-1}{2}
\hat {\bf p}1 \times (\hat {\bf p}1 \times \hat {\bf q}1)] - \right.
\nonumber\\*
&&\left. \frac{\cos (\varphi (p1))}{p1} (\vec 
\Sigma_{s3s1}^{*} \times
\vec \sigma_{s1''s1'}) \cdot \hat {\bf q}1 \right]\;
(-\frac{1}{\sqrt{3}})
\nonumber\\*
&&\nonumber\\*
&& terms [-]: (N\;\pi )_{11}^{-} 
\nonumber\\*
&&2 [+\frac{i}{2} \frac{1-\sin (\varphi( p1))}{p1}] 
\left[ \frac{1}{2}
[\varphi' (q1)+ \frac{\cos (\varphi (q1))}{q1} (
\vec \Sigma_{s1''s1'} \times \vec \sigma_{s3s1} )\cdot [\frac{-1}{2}
\hat {\bf q}1 \times (\hat {\bf q}1 \times \hat {\bf p}1)] - \right.
\nonumber\\*
&&\left. \frac{\cos (\varphi (q1))}{q1} (\vec 
\Sigma_{s1''s1'} \times
\vec \sigma_{s3s1}) \cdot \hat {\bf p}1 \right]\;
(-\frac{1}{\sqrt{3}}) 
\end{eqnarray}
In $term\; b)$ the contributions of ${(n \pi )_{12}}$ and ${(n \pi )_{13}}$
are zero. This is not so in the case of the $n\pi \Delta $ coupling where we 
will have an extra contribution coming from the sum of the diagrams 
$(N\;\pi )_{12}^{-}$\ and  
$(N\;\pi )_{12}^{+}$ which survive the cancelation mechanism of Eq. 
(\ref{corsat}) and which will contribute along the same lines as before to a 
new overlap ${\cal O}'$, 
\begin{eqnarray}\label{casinhos}
&&{\bf Term\;b'+})\nonumber\\*
&&\left[-\frac{1}{2}(\varphi ' (p1)+\frac{C1}{p1}) \hat {\bf p}1 \times 
(\hat {\bf p}1 \times \hat {\bf p}2)
-\frac{C1}{p1} \hat {\bf p}2\right] [\frac{1-S2}{2p2}]\cdot 
(i \vec \sigma _{qk1} 
\times \vec \sigma _{qk2})\; (-\frac{1}{\sqrt{3}})
\nonumber\\*
&&{\bf Term\;b'-})\nonumber\\* 
&&\left[-\frac{1}{2}(\varphi ' (q1)+\frac{C1}{q1}) \hat {\bf q}1 \times 
(\hat {\bf q}1 \times \hat {\bf q}2)
-\frac{C1}{q1} \hat {\bf q}2\right] [\frac{1-S2}{2q2}]\cdot 
(i \vec \sigma _{qk1} 
\times \vec \sigma _{qk2})\; (-\frac{1}{\sqrt{3}})
\end{eqnarray}
where $C1$ and $S2$ stand respectively for $\cos (\varphi (p1))$ and 
$\sin (\varphi (p2)) $ for $Term\;b'+)$ and $\cos (\varphi (q1))$ and 
$\sin (\varphi (q2)) $ for $Term\;b'-) $. In expression (\ref{casinhos} 
as well as in the remainder of this paper $\sigma _{qk1}$ represents 
the Pauli matrix operator acting in quark 1. A similar notation is used 
for quark 2.

In order to obtain the results of Eq. (\ref{casos}, \ref{casinhos}) we have 
made use of the following identity, 
\begin{equation}
( \vec \Sigma_{s3s1}^{*} \cdot \hat {\bf p})\; (\hat {\bf p} \cdot
[ \vec \sigma_{s1''s1'} \times \hat {\bf q} ] ) =
-\frac{1}{2} ( \vec \Sigma_{s3s1}^{*} \times \vec \sigma_{s1''s1'} \cdot
(\hat {\bf p} \times [ \hat {\bf p} \times \hat {\bf q} ] )
\end{equation}
which is valid both for the nucleon and delta cases because the matrix 
elements of the tensor operator $\langle N |T_2 (\vec \sigma_1,\vec \sigma_2 )
\cdot T_2({\bf p1},\;{\bf p2})|nN\rangle $ are zero.

Finally we have also to consider the spin  wave function of the pion,
$i [ \sigma_{2}]_{s1',s3}/\sqrt{2}$, and using the simple relations,
\begin{eqnarray}                   
&&(i\; \sigma_{2}\; )\vec \Sigma^{*}= \vec \sigma 
\nonumber\\*
&&\vec \Sigma  (i\sigma_{2 })= -\vec \sigma 
\end{eqnarray}
we are able to write, 
\begin{eqnarray}\label{casos1}
&& Term\;a\  +)\; \nonumber\\*
&&\{ \varphi '(p1) \vec \sigma \cdot \nabla N_{out}+
(\varphi '(p1) +\frac{\cos (\varphi (p1))}{p1}) \vec \sigma \cdot
\hat {\bf p}1 \times (\hat {\bf p}1 \times \nabla N_{out} )\} 
\Phi_{s}^+ N_{in} 
\nonumber\\*
&& Term\;a\  -)\; \nonumber\\*
&&N_{out}\Phi_{s}^{-}\{ \varphi '(q1) \vec \sigma \cdot \nabla N_{out}+
(\varphi '(q1) +\frac{\cos (\varphi (q1))}{q1}) \vec \sigma \cdot
\hat {\bf q}1 \times (\hat {\bf q}1 \times  \nabla N_{in} )\} 
\nonumber\\*
&& Term\;b\ +)\; \nonumber\\* 
&& \frac{1-\sin (\varphi (q1))}{2 q1} \{ 2 \frac{\cos (\varphi (p1))}{p1} 
\vec \sigma \cdot \hat {\bf q}1+
(\varphi '(p1) +\frac{\cos (\varphi (p1))}{p1}) \vec \sigma \cdot
\hat {\bf p}1 \times (\hat {\bf p}1 \times \hat {\bf q}1 ) \} N_{out}  
\Phi_{s}^+ N_{in}
\nonumber\\*
&& Term\;b\ -)\; \nonumber\\* 
&&-\frac{1-\sin (\varphi (p1))}{2 p1} \{ 2 \frac{\cos (\varphi (q1))}{q1} 
\vec \sigma \cdot \hat {\bf p}1+
(\varphi '(q1) +\frac{\cos (\varphi (q1))}{q1}) \vec \sigma \cdot
\hat {\bf q}1 \times (\hat {\bf q}1 \times \hat {\bf p}1) \}  N_{out} 
\Phi_{s}^+ N_{in},
\nonumber\\*
\end{eqnarray}
where $\Phi_{s}$ represents the pion Salpeter amplitude amputated of its 
spin wave function factor and we have omitted the factor 
$(-\frac{1}{\sqrt{3}})$ of the color overlap.

At this stage it is convenient to introduce the 
following vector $\vec v_{N-\pi -N}$, summarizing the spin-flavor overlap 
both for the $n\pi n$ and $n\pi \Delta$ cases obtained in appendix C (Eq. 
\ref{spinflave}),
\begin{eqnarray}
\vec v_{n \pi n }&&=\frac{1}{2}\{ \frac{5}{9},\; 0 \}\nonumber\\*
\vec v_{n \pi \Delta }&&=\frac{1}{2}\{ \frac{2\;\sqrt{2}}{3},\; \sqrt{2} \}
\end{eqnarray}
The factor $\frac{1}{2}$ stands for the spin-flavor overall normalization of 
the pion wave function.
The next step is to expand ${\cal O}$ and ${\cal O}'$ up 
to first order in P and to integrate both in ${\bf k}''$ and in the 
solid angle of ${\bf k}$. It is also convenient to introduce the vector 
$\vec {\cal T}=\{ {\cal O}_{fs},\; {\cal O}'_{fs} \}$, which correspond to 
the generic $N\pi N$ overlaps with both the spin-flavor and  
$\vec \sigma_{n} \cdot {\bf P}\ \vec \tau_{N}$ (for the $n\pi n$ case) or 
$ \vec S \cdot {\bf P}\; \vec T$ (for the $n\pi \Delta$ case) 
factors amputated. We have,
\begin{eqnarray}\label{sfamputated1}
{\cal O}_{fs}&&=\sqrt{3}\;
\frac{\int k^2 dk [{\cal C} \sin (\varphi )+ {\cal G}]\  
e^{-\frac{3 k^2}{2 \alpha^2}}}{\int k^2 dk 
e^{-\frac{3 k^2}{2 \alpha^2}}}
\nonumber\\*
{\cal C}&&=\{ -\frac{2}{\alpha^2} 
[\frac{\varphi '}{6}-
\frac{k^2 \sin (\varphi )}{2}+\frac{3 \sin (\varphi )\cos (\varphi )-
5 \cos (\varphi )}{6 k} ]-\frac{(1-\sin (\varphi )) \varphi '}{k^2}\}
\nonumber\\*
{\cal G}&&=\{-\frac{2}{\alpha^2} \}
\frac{k \varphi '}{2} g_1 
\end{eqnarray}
and we use the approximate relations,
\begin{equation}
\frac{\cos \varphi }{k}\simeq \frac{1-\sin \varphi }{k^2}\simeq 
2.05\; e^{-\frac{2k^2}{\alpha^2}}
\end{equation}
valid for the chiral angle solution of the mass gap equation (\ref{massgap}) 
depicted in figure \ref{fchir} to obtain,
\begin{eqnarray}\label{sfamputated2}
{\cal O}_{fs}'&&=\sqrt{3}\;
\frac{\int k^2 dk [{\cal D}\; \sin (\varphi (k))]\ 
e^{-\frac{3 k^2}{2 \alpha^2}}}{\int k^2 dk 
e^{-\frac{3 k^2}{2 \alpha^2}}}  
\nonumber\\*
{\cal D}&&\simeq 
-\frac{1.4}{2\sqrt{2}} \left[ \frac{1}{2}+\frac{5k^2}{4\alpha^2}\right]
\; e^{-\frac{9k^2}{4\alpha^2}}
\end{eqnarray}
so that the final overlap is given by,
\begin{equation}
f_{N-\pi -N}=\frac{m_{\pi }}{a}\;\vec v_{N-\pi -N} \cdot \vec {\cal T}
\end{equation}

\section{Color Overlaps}

When performing color calculations we have to attend to the following rules:
\begin{eqnarray}
&&I-\; One\; quark\nonumber\\*
&& \frac{\lambda }{2}\cdot \frac{\lambda }{2}=\frac{4}{3}\nonumber\\*
&&II-\; One\; Nucleon\;(1,2,3)\nonumber\\*
&&(\lambda_{1}+\lambda_{2} +\lambda_{3})^2=0\rightarrow
{\left( \frac{\lambda }{2}\right) }_{i}\cdot
{\left( \frac{\lambda }{2}\right) }_{j}=
-\frac{2}{3}\hspace{2mm} |_{i,j=1,2,3}
\nonumber\\*
&&III-\; One Meson\ q-\overline{q}\nonumber\\*
&&\frac{\lambda_{q}}{2}\cdot \frac{\lambda_{\overline{q}}}{2}=
-\frac{4}{3}\nonumber\\*
&&IV-\; quark-\lambda -\overline{quark}\; vertex\nonumber\\* 
&&\overline{q}_{\alpha} \;\lambda\;q_{\beta}=\overline{q}_{\alpha}\;
(q_{\alpha}\lambda\;q_{\beta})=
-(\overline{q}_{\alpha}\;\lambda\;\overline{q}_{\beta})\; q_{\beta}\nonumber\\*
&&V-\; quark\; exchange\nonumber\\*
&&\langle N(1,2,3)\;  M(4,{\overline{5}}) |{\cal P}^{i4} |
N(1,2,3)\; M(4,\;\overline{5})\rangle =\frac{1}{3}
\end{eqnarray}
so that we have,
\begin{eqnarray}
&&\langle N |(-\frac{3}{4})\frac{\lambda }{2}\cdot \frac{\lambda }{2} |
N \rangle =\frac{1}{2}\nonumber\\*
&&\langle M |(-\frac{3}{4})\frac{\lambda }{2}\cdot \frac{\lambda }{2} |
M \rangle =1
\end{eqnarray}
In the table below we summarize the color matrix elements for the
$N-\pi-N$ coupling ${\cal O}$.
\begin{quasitable}
\begin{tabular}{lclc}
$Diagram$&$-\frac{3}{4} \frac{\lambda }{2} \cdot \frac{\lambda }{2}
$&$Diagram$&$-\frac{3}{4} \frac{\lambda }{2} \cdot \frac{\lambda }{2}$\\
\tableline
$ (N\pi )_{11}^{+}$&$\frac{-1}{\sqrt{3}}$&
$ (N\pi )_{11}^{-}$&$\frac{-1}{\sqrt{3}}$\\
$ (N\pi )_{12}^{+}$&$\frac{1}{2 \sqrt{3}}$&
$ (N\pi )_{12}^{-}$&$\frac{1}{2\sqrt{3}}$\\
$ (N\pi )_{13}^{+}$&$\frac{1}{2 \sqrt{3}}$&
$ (N\pi )_{13}^{-}$&$\frac{1}{2\sqrt{3}}$\\
\end{tabular}
\end{quasitable}

\section{Spin and Flavor Overlaps}

The spin-flavor content of the nucleon and delta wave functions is given by,
\begin{equation}\label{wavespin}
|n\rangle =\frac{1}{\sqrt{2}} |D_{spin} D_{flavor}
+ F_{spin} F_{flavor}\rangle.
\end{equation}
It is sufficient to workout the spin overlaps being the flavor overlaps
identical. The spin D wave functions are given by,
\begin{eqnarray}\label{nucflasp}
|n\uparrow \rangle &=&\sqrt{\frac{2}{3}}|1\;1,\;\frac{1}{2}\;
-\frac{1}{2}\rangle -
\sqrt{\frac{1}{3}}|1\;0,\;\frac{1}{2}\;\frac{1}{2}\rangle \nonumber\\*
|n\downarrow \rangle &=&-\sqrt{\frac{2}{3}}|1\;-1,\;\frac{1}{2}\;
\frac{1}{2}\rangle +
\sqrt{\frac{1}{3}}|1\;0,\;\frac{1}{2}\;-\frac{1}{2}\rangle 
\end{eqnarray}
where the first two numbers of each ket represent the spin content of
the quark pair 1-2. Similarly we have for the F wave functions,
\begin{eqnarray}\label{nucFfs}
|n\uparrow \rangle &=& |0\;0,\;\frac{1}{2}\;\frac{1}{2}\rangle \nonumber\\*
|n\downarrow \rangle &=& |0\;0,\;\frac{1}{2}\;-\frac{1}{2} \rangle
\end{eqnarray}
For the delta we have,
\begin{equation}\label{DELTA}
|\Delta \rangle =|{3 \over 2} s\rangle_{spin} \otimes |{3 \over 2} f\rangle_{flavor}.  
\end{equation}\label{spiflavpi}
The pion spin-flavor wave function is given by,
\begin{equation}\label{spin-fla}
\chi_{\pi }=\{\frac{i}{\sqrt{2}} \sigma_{2}\}_{spin} \otimes
\{ \frac{1}{\sqrt{2}} \sigma_{3},\sigma_{\pm}\}_{flavor}
\end{equation}
where $\sigma_{i} $ stands for the appropriate Pauli matrix.
$\sigma_{\pm}=(\sigma_{1}\pm i \sigma_{2})/2$. From now on we
will follow the usual notation in what concerns isospin and denote
$\sigma_{flavor}=\tau $.

For the $N\pi N$ overlaps  we need to consider the expectation value of 
the operators $\sigma_{qk1}$ and $\sigma_{qk1}\otimes \sigma_{qk2}$. We 
calculate a few examples to illustrate the general method of calculating these 
operator matrix elements. The other cases are obtained in the same manner.
A general matrix element looks like,
\begin{equation}
\langle B^\dagger |\vec \sigma_{qk1} \otimes \vec \sigma_{qk2} |A \rangle =
trac\{ B^\dagger \vec \sigma_{qk1} \otimes A \vec \sigma^{t}_{qk2} \}
\end{equation}
where the generic form for the spin wave functions B and A of two 
quarks are given by,
\begin{equation}
A=(a_0 + \vec a_1 \cdot \vec \sigma ) (i \sigma_2)
\ B=(b_0 + \vec b_1 \cdot \vec \sigma ) (i \sigma_2)
\end{equation}
The pion spin wave function is for instance $a_0 (i \sigma_2)$ whereas 
the vector $\rho$ is given by $(\vec b_1 \cdot \vec \sigma ) (i \sigma_2)$.  
It is also convenient to write the spin-flavor content of the nucleon 
wave function as follows,
\begin{equation}
\displaystyle
(n \uparrow,\; n\downarrow\; )_{\;spin\;nuc}=(\uparrow,\; \downarrow)_{qk3}\ 
\left( \frac{1}{\sqrt{2}}\;\frac{F_f}{\sqrt{2}}\ 1_{(qk3,\;spin\;nuc)}-
\frac{1}{\sqrt{6}}\;\frac{D_f}{\sqrt{2}}\ \sigma_{(qk3,\;spin\;nuc)}
\right)
\end{equation}
which maps the spin content of the nucleon to the spin of the third quark. 
Next we use the following identities,
\begin{eqnarray}
&&trac \{ \sigma_i \sigma_j \} = 2 \delta_{ij} \nonumber\\*
&&trac \{ \sigma_i \sigma_b \sigma_j \} =2 i \epsilon _{ibj} \nonumber\\*
&&trac \{ \sigma_i \sigma_b \sigma_j \sigma_a \} = 
2( \delta _{ib} \delta _{ja} - \delta _{ij} \delta _{ba} + \delta _{ia} 
\delta _{jb} ),
\end{eqnarray}
to obtain,
\begin{eqnarray}
\displaystyle
\langle B^\dagger |\sigma_{qk1_i}  \sigma_{qk2_j} |A \rangle &&= 
-2 b_0^\dagger  a_0 \delta_{ij}+2 i {b_0}^\dagger a_{1_k} \epsilon_{ijk}
\nonumber\\*
&&- 2 i a_0 {b_{1_k}}^\dagger \epsilon_{ijk}-2 {b_{1_i}}^\dagger a_{1_j} -
2 {b_{1_j}}^\dagger a_{1_i}+2 \vec b_1 \cdot \vec a_1 \delta_{ij},
\end{eqnarray}

We have to consider three cases of tensors,
\begin{eqnarray}
&&\langle B^\dagger |\vec \sigma_{qk1} \cdot \vec \sigma_{qk2}\ |A \rangle =
-6 b_0^\dagger \; a_0 +2\;\vec b_1 \cdot \vec a_1\nonumber\\*  
&&\langle B^\dagger |i\; \vec \sigma_{qk1} \times \vec \sigma_{qk2} 
|A \rangle =
-4\; (b_0^\dagger\; \vec a_1 -a_0 \vec b_1^\dagger )
\end{eqnarray}

Then it is simple to evaluate for instance,
\begin{eqnarray}
&&\langle F_s |\vec \sigma_{qk1} |D_{s'}\rangle =2\; \frac{1}{\sqrt{2}}
\; \frac{-1}{\sqrt{6}} \vec \sigma_{s,\; s'} \nonumber\\*
&&\langle D_s |\vec \sigma_{qk1}   |D_{s'}\rangle=-2i\; 
\frac{-1}{\sqrt{6}}\; \frac{-1}{\sqrt{6}}\; 
(\vec \sigma_{qk1} \times \vec \sigma_{qk2} ) =\frac{2}{3}\sigma_{s,\; s'}
\end{eqnarray}
Similar calculations for the other matrix elements so that for the 
$n\pi n$ overlap we have the following spin matrix elements,
\begin{quasitable}
\begin{tabular}{llclc}
&$\langle D\;D\rangle$&$\langle F\;D\rangle$&$\langle D\;F\rangle$&
$\langle F\;F\rangle$\\
\tableline
$\vec \sigma_{qk1}$&$\frac{2}{3}\vec \sigma_{n}$&$0$&
$-\frac{1}{\sqrt{3}}\vec \sigma_{n}$&$-\frac{1}{\sqrt{3}}\vec \sigma_{n}$\\
$i\vec \sigma_{qk1}\times \vec \sigma_{qk2}$&$0$&$0$&$-\frac{2}{\sqrt{3}} \vec 
\sigma_{n}$&
$\frac{2}{\sqrt{3}} \vec \sigma_{n}$
\end{tabular}
\end{quasitable}
where $\sigma_{n}$ represents the Pauli matrix acting in spin of the 
nucleon as a whole. We can repeat this calculations in the flavor space, using 
the appropriate flavor representation for the pion wave function given in 
 equation (\ref{spin-fla}) so that  we have for the $n \ n$ spin-flavor 
overlap,
\begin{equation}\label{spinfn}
\langle n |
\{ \vec \sigma_{qk1},\;i\vec \sigma_{qk1}
\times \vec \sigma_{qk2} \}\otimes \vec \tau_{qk1} | n \rangle
= \{ \frac{5}{9} \vec \sigma_{n} \otimes \vec \tau_{n},
0\}
\end{equation}

For the $n \pi \Delta$ overlap we will need the following spin matrix
elements,
\begin{quasitable}
\begin{tabular}{llc}
&$\langle \Delta\;D\rangle$&$\langle \Delta\;F\rangle$\\
\tableline
$\vec \sigma_{qk1}$&$\frac{1}{\sqrt{3}}\vec S$&$ \vec S$
\\
$i\vec \sigma_{qk1}\times \vec \sigma_{qk2}$&$0$&$2 \vec S$
\end{tabular}
\end{quasitable}
Again we repeat similar calculations, this time in the flavor space,
so that we have for the $\Delta \ n$ spin-flavor overlap,
\begin{equation}\label{fsdelta}
\langle \Delta |
\{ \vec \sigma_{qk1},\;i\vec \sigma_{qk1}
\times \vec \sigma_{qk2} \}\otimes \vec \tau_{qk1} | n \rangle
= \{ \frac{2 \sqrt{2}}{3} \vec S \otimes \vec T,
\sqrt{2}\ \vec S \otimes \vec T \}
\end{equation}
where $\vec S$ and $\vec T $ are given in ref.\cite{Ericson} and are 
defined by, 
\begin{equation}
\langle \frac{3}{2} \nu _{\Delta}|(S^{\dagger}_{\lambda},\;
T^{\dagger}_{\lambda})|\frac{1}{2}\; \nu _{N} \rangle =\langle 
\frac{3}{2} \nu _{\Delta}|1\; \lambda\ \frac{1}{2} \nu_{N} \rangle
\end{equation}

Finally from Eqs. (\ref {spinfn}, \ref{fsdelta}) we can construct the vectors 
$\vec v_{N-\pi -N}$,
\begin{eqnarray}\label{spinflave}
\vec v_{n \pi n }&&=\frac{1}{2}\{ \frac{5}{9},\; 0 \}\nonumber\\*
\vec v_{n \pi \Delta }&&=\frac{1}{2}\{ \frac{2\;\sqrt{2}}{3},\; \sqrt{2} \}
\end{eqnarray}

\begin{figure}
\caption{\label{fchir}The chiral angle as a function of k}
\caption{\label{ffeyn}Feynman rules in the spin representation}
\caption{\label{fDyson}Dyson series for $q-\bar q $ bound states}
\caption{\label{f1g1}The $f_1\ and\ g_1$ amplitudes contributing to the 
boosted pion}
\caption{\label{fNucl}The salpeter equation for the nucleon and 
the $\Delta$}
\caption{\label{foverl}Diagrams contributing to the $N \pi N$ coupling}
\end{figure}

\begin{references}
\bibitem{Ribeiro} M. Oka and K. Yazaki Phys Lett. {\bf 90}, 41 (1980); 
J. Ribeiro, Zeit. Phys. C5, 27 (1980)
\bibitem{KN} P.Bicudo, J. Ribeiro , J.Rodrigues, Phys. Rev. C 
{\bf 52}, 2144 (1995)
\bibitem{BicRib1} P. Bicudo and J. Ribeiro, Phys.
Rev. D {\bf 42}, 1635 (1990).
\bibitem{BicKre}P. Bicudo, G. Krein, J. Ribeiro and J. Villate, Phys. Rev. D
{\bf 45}, 1673 (1992)
\bibitem{VilLiu} J.Villate, D.S. Liu, J.E. Ribeiro, P.Bicudo, Phys. Rev. D
{\bf 47}, 1145 (1993)
\bibitem{Skyrme} J. M. Eisenberg in V Hadron Physics 1996, edited by
 Erasmo Ferreira (to be published in Word Scientific)
\bibitem{NJL} Y. Nambu, G. Jona-Lasinio, Phys. Rev. {\bf 122}, 345 (1961);
{\bf 124}, 246 (1961).
\bibitem{AmeYao} A. Amer, A. Le Yaouanc, L. Oliver, O. P\`ene and J-C. Raynal,
Particles and Fields, {\bf 17},
61 (1983); A. Le Yaouanc, L. Oliver, O. P\`ene and J-C. Raynal, Phys. Rev. D
{\bf 29}, 1233 (1984); A. Le Yaouanc, L. Oliver, S. Ono, O. P\`ene and
J-C. Raynal, {\em ibid.} {\bf 31}, 137 (1985).
\bibitem{BicRib2} P. Bicudo and J. Ribeiro, 
Phys. Rev. D {\bf 42}, 1611 (1990);  
\bibitem{BicRib3}P. Bicudo and J. Ribeiro, Phys. Rev. D {\bf 42}, (1990) 1625.
\bibitem{AdlDav} S. Adler and A. C. Davis, Nuc. Phys. B {\bf 244}, 469 (1984); 
\bibitem{Covar} 
H. Pagels Phys Rev D {\bf 14} 2747 (1976); 
H. Pagels Phys Rev D {\bf 15}, 2991 (1977); 
Y. Dai, Z. Huang and D. Liu, Phys. Rev. D{\bf 43}, 1717 (1991).
%
\bibitem{Horvat}
Y. L. Kalinovsky, L. Kaschlun and V. N. Pervushin, Phys. Lett.
B, {\bf 231}, 288 (1989); Y. L. Kalinovsky, W. Kallies, B. N. Kuranov, 
V. N. Pervushin and N. A. Sarikov, Yad. Fiz. {\bf 49}, 1059 (1989);
V. N. Pervushin, Y. L. Kalinovsky, W. Kallies, N. A. Sarikov, Fortschr. Phys. 
{\bf 38}, 333 (1989); 
R. Horvat, D. Kekez,D. Palle, 
D. Klabucar, Zagreb Preprint {\bf ZTF-93-9-R}
(1993);
R. Horvat, D. Kekez, D. Klabucar, D. Palle Phys Rev D {\bf 44}, 1584 (1991).
%
\bibitem{Spit} W. F. M. Spit, A.G.M. van Hees, P. J. Brussaard and 
P. J. Mulders, Nuc. Phys. A {\bf 570}, 472 (1994)
%
\bibitem{nucmat} P. Bicudo Phys Rev Lett {\bf 72}, 1600 (1994) 
%
\bibitem{Brambilla}N Brambilla, P. Consoli and G. M. Prosperi, 
Phys Rev D {\bf 50}, 5878 (1994)
%
\bibitem{Ericson} T.Ericson, W. Weise in $Pions\ and\ Nuclei$ 
(Oxford Science Publications, 1988)
\bibitem{QPCMM} A. le Yaouanc, L. Oliver, O. Pene and J.-C. Raynal,
Phys Rev D {\bf 8}, 2223 (1973)
%
\bibitem{GraphicalRules} J.E. Ribeiro, Phys Rev D {\bf 25}, 2406 
(1982)
\end{references}
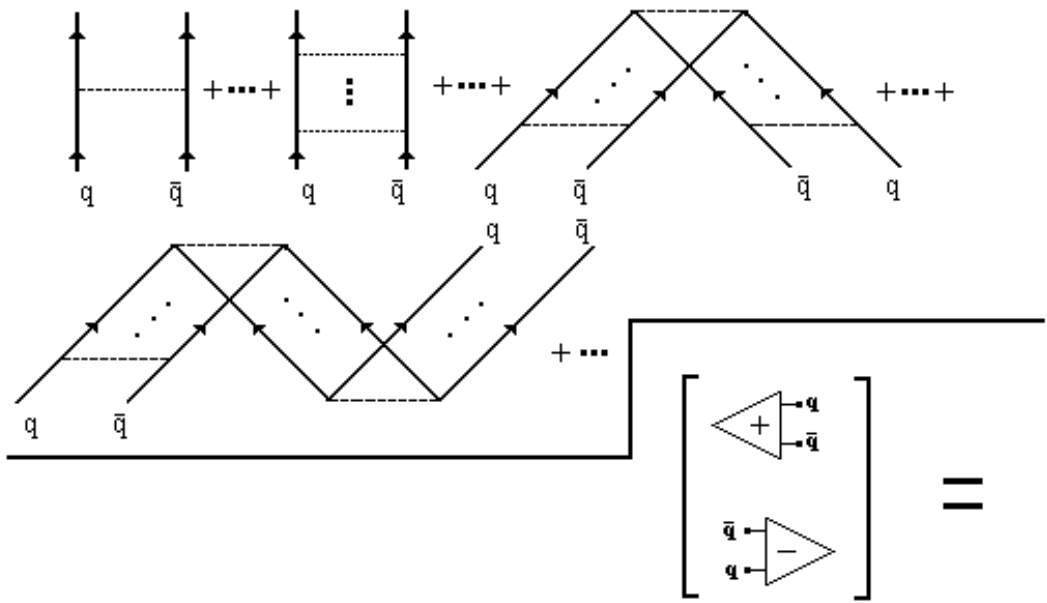
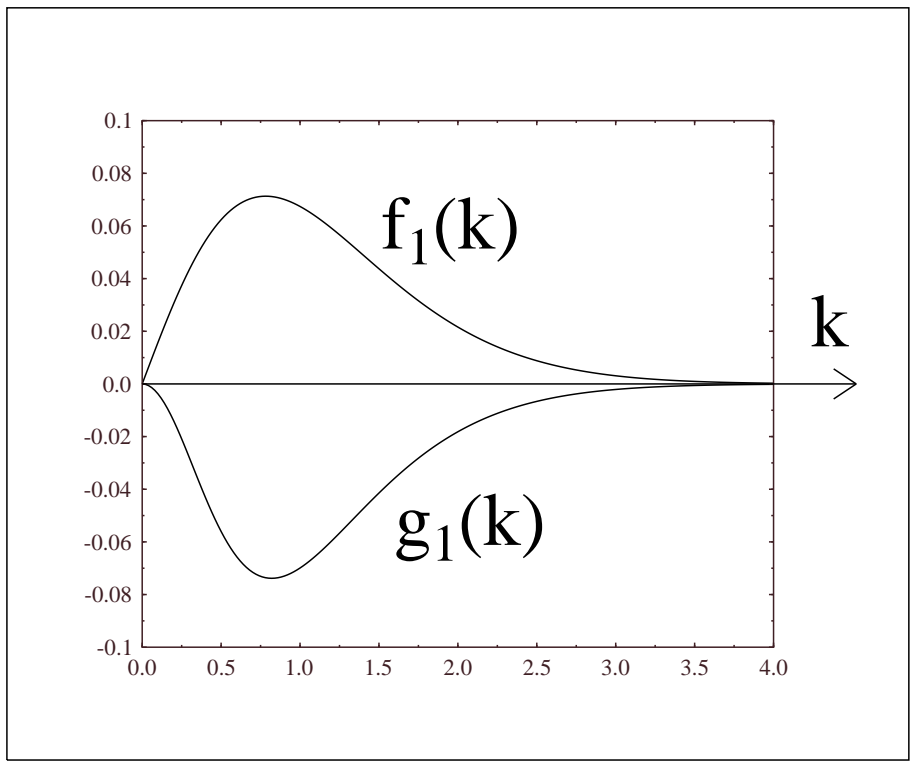
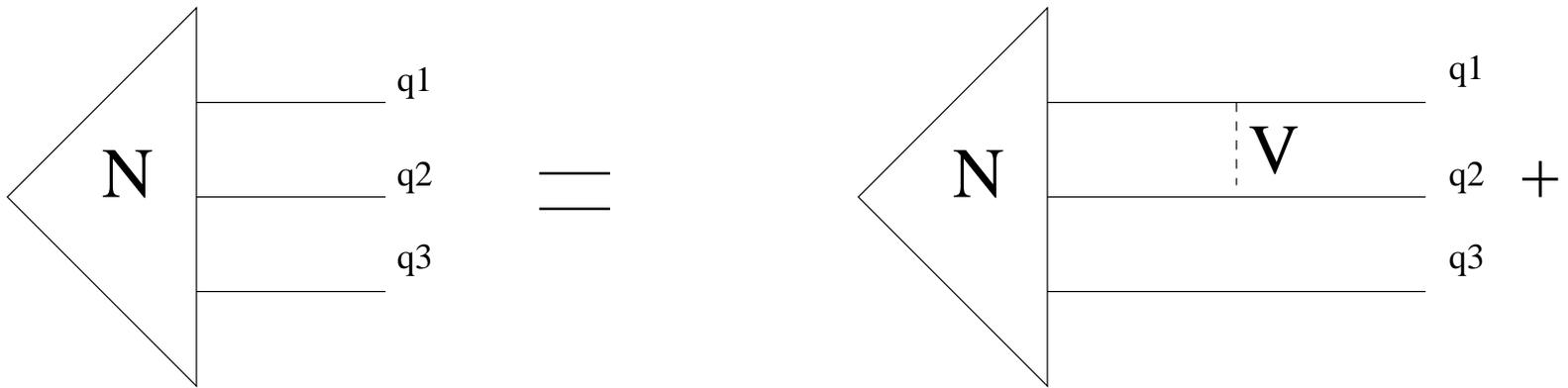
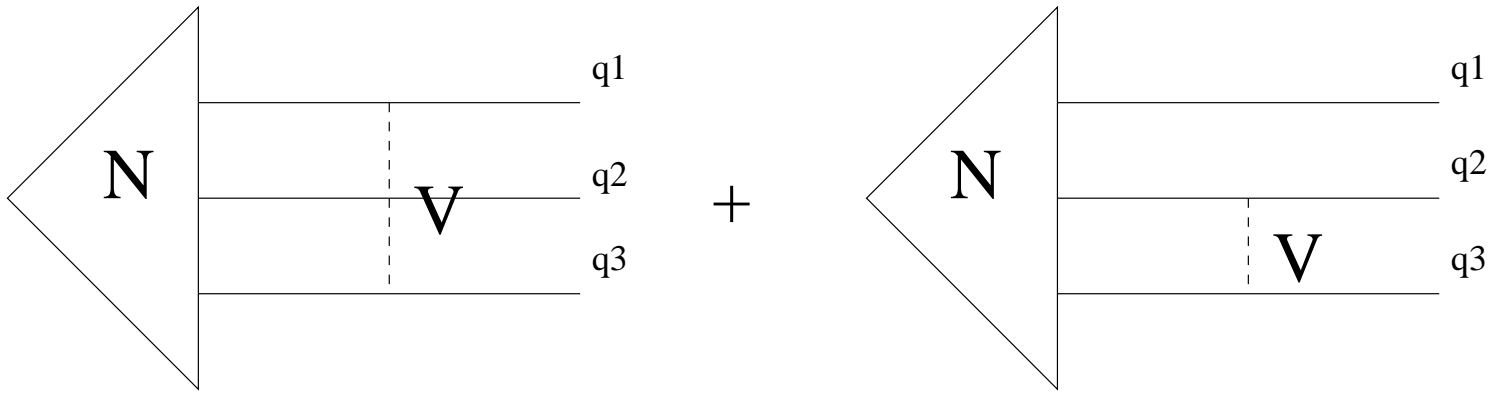
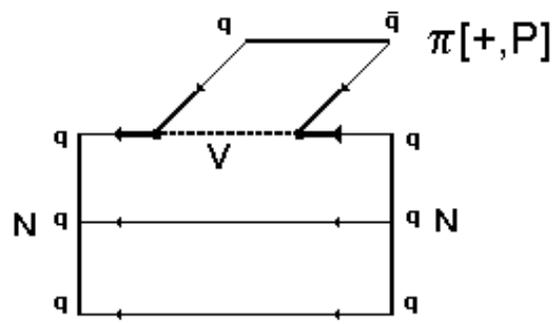
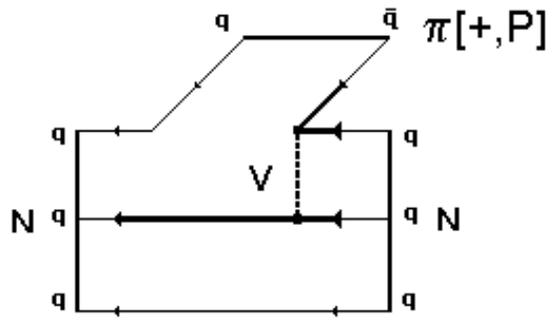
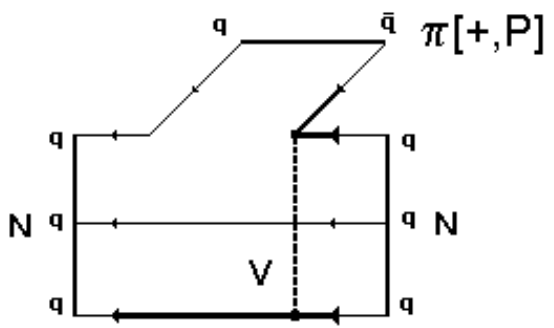
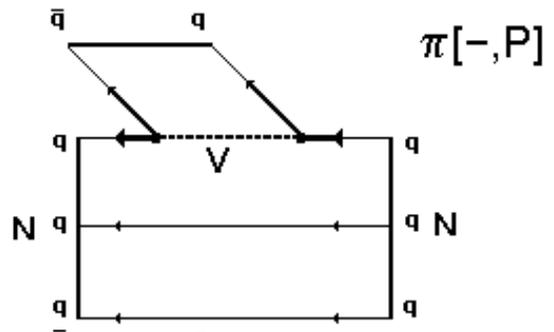
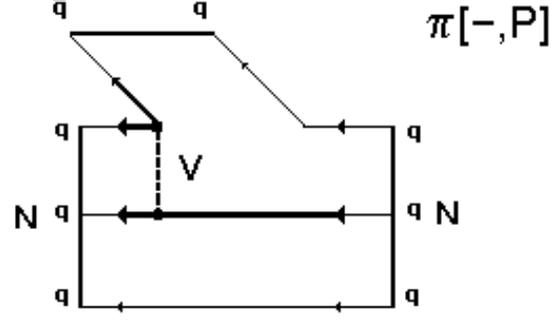
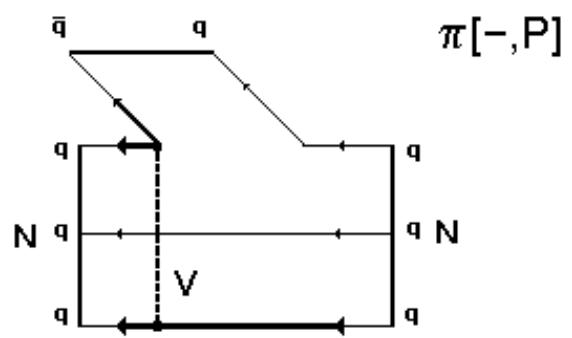
\end{document}